\documentclass[conference, a4paper]{IEEEtran}
\IEEEoverridecommandlockouts
\usepackage{cite}

\usepackage{romannum}
\usepackage{mathtools}
\DeclarePairedDelimiter{\ceil}{\lceil}{\rceil}

\ifCLASSINFOpdf
\usepackage[pdftex]{graphicx}
\graphicspath{{../pdf/}{../jpeg/}{./png/}}
\DeclareGraphicsExtensions{.pdf,.jpeg,.png}
\else
  \usepackage[dvips]{graphicx}
  \graphicspath{{Figures/eps/}}
\DeclareGraphicsExtensions{.eps}
\fi

\usepackage{amsmath}
\usepackage{stfloats}

\usepackage{color}

\usepackage{balance}

\begin{document}
%
\title{Secret Key Generation Rates for Line of Sight Multipath Channels in the Presence of Eavesdroppers}


\author{
    \IEEEauthorblockN{Amitha Mayya$^\star$, Arsenia Chorti$^{\star,\dagger}$, Rafael F. Schaefer$^{\star,\ddagger}$, Gerhard P. Fettweis$^{\star,\ddagger}$\\[1ex]}
    \IEEEauthorblockA{$^\star$ Barkhausen Institut, Dresden, Germany \\
    $^\dagger$ ETIS UMR 8051, CYU, ENSEA, CNRS, Cergy, France\\
    $^\ddagger$ Technische Universit\"at Dresden, Germany \& BMBF Research Hub 6G-life \& Cluster of Excellence CeTI \\[0.75ex]
    \{amitha.mayya, arsenia.chorti, rafael.schaefer, gerhard.fettweis\}@barkhauseninstitut.org}
    \thanks{This publication is supported based upon work from COST Action 6G-PHYSEC (CA22168), supported in part by COST (European Cooperation in Science and Technology) and in part by the Saxon State government out of the State budget approved by the Saxon State Parliament. The work of A. Chorti is further supported in part by the EC through the Horizon Europe/JU SNS project ROBUST-6G (Grant Agreement no. 101139068), the ANR-PEPR 5G Future Networks project, the ELIOT ANR-18-CE40-0030 and FAPESP 2018/12579-7, FAPERJ project and the INEX-PHEBE project. The work of R. F. Schaefer and G. P. Fettweis is further supported in part by the German Federal Ministry of Education and Research (BMBF) within the National Initiative on 6G Communication Systems through the Research Hub \emph{6G-life} under Grant 16KISK001K and in part by the German Research Foundation (DFG) as part of Germany’s Excellence Strategy – EXC 2050/1 – Project ID 390696704 – Cluster of Excellence \emph{``Centre for Tactile Internet with Human-in-the-Loop'' (CeTI)} of Technische Universit\"at Dresden.}
}


%


\maketitle

\begin{abstract}

In this paper, the feasibility of implementing a lightweight key distribution scheme using physical layer security for secret key generation (SKG) is explored. Specifically, we focus on examining SKG with the received signal strength (RSS) serving as the primary source of shared randomness. Our investigation centers on a frequency-selective line-of-sight (LoS) multipath channel, with a particular emphasis on assessing SKG rates derived from the distributions of RSS. We derive the received signal distributions based on how the multipath components resolve at the receiver. The mutual information (MI) is evaluated based on LoS 3GPP channel models using a numerical estimator. We study how the bandwidth, delay spread, and Rician $K$-factor impact the estimated MI. This MI then serves as a benchmark setting bounds for the SKG rates in our exploration.
\end{abstract}


%
\IEEEpeerreviewmaketitle

\section{Introduction}
The prevailing security issues in 5G systems such as false base station attacks, jamming attacks, and the fact that the current standard public key encryption algorithms are not quantum secure, raise further security concerns that need to be addressed in 6G. In addition to this, security attacks can take the form of denial-of-service attacks and passive eavesdropping attacks. The machine type communications in the industrial internet of things (IIoT) domain, which includes complex interconnected and automated networks of embedded machines and sensors, create new vulnerabilities where cryptographic security solutions cannot be implemented due to the low computation power or insufficient memory capabilities. Hence, in IIoT there is a pressing need for resilient and trustworthy security solutions that should be of low complexity in addition to being quantum secure\cite{jsac,chorti2022contextaware,plsi}. 

Physical layer security (PLS) leverages the unique characteristics of wireless communication channels to enhance the confidentiality and integrity of data transmission \cite{BlochBarros-2011-PhysicalLayerSecurity,SchaeferBocheKhistiPoor-2017-InformationTheoreticSecurityPrivacy,pnas}. Rather than relying solely on cryptographic algorithms and higher layer security protocols, PLS takes advantage of the inherent properties of the wireless channel which are noisy and imperfect in nature and implements the security solutions in the physical layer. PLS makes it possible to develop security solutions that are lower in complexity and hence can be used in IoT devices which are constrained by complexity, memory, and computational power. PLS can also be used as a complementary solution to the existing cryptographic solutions and thereby enhance the overall security of the communication system.

PLS-based secret key generation (SKG) enables two nodes to extract a shared secret \cite{Maurer,ahlswede}. This protocol involves four steps: randomness extraction, quantization, information reconciliation, and privacy amplification. In the randomness extraction phase, legitimate users extract correlated channel observations, which are then converted into bits during the quantization step. Mismatches arising from receiver noise and imperfect channel estimations are corrected in the reconciliation phase using distributed source coding techniques. Evaluation of potential leakage occurs during the privacy amplification stage, ensuring the generation of secure confidential keys.



When waveforms are transmitted, the waveform undergoes attenuation and shaping due to the small scale and large scale propagation effects of the environment. The small scale fading due to the multipath of the fading channel and Doppler spread can be modelled as stochastic random process. But due to the reciprocity theorem \cite{reciprocity_Smith}, both the antennas at the legitimate users observe identical random fading effects and this is used as a common source of randomness from which secret keys can be extracted during the channel coherence time. Since the wireless channel measured at two different instances of time will be uncorrelated (provided they are greater than the channel coherence time), this will be considered as the source of shared randomness. 

In the realm of SKG literature, there is a commonly held belief, according to Jake's model, that the channel undergoes decorrelation at a distance equal to half a wavelength \cite{half_wavelength}, \cite{Goldsmith}. It's worth noting, though, that this assumption remains valid solely under the condition of an environment featuring infinite uniformly distributed scattering \cite{He2016linksignature}. Therefore, in real-world scenarios, it becomes crucial to consider and incorporate the correlations between the observations of legitimate nodes and potential eavesdroppers.

If the received channel observations (which are considered as source of shared randomness to distill the secret key) at the two antennas (legitimate users Alice and Bob) are considered as a sequence of $N$ random variables (RVs) given by, $A_N = [A_1, ..., A_N]$, $B_N = [B_1, ..., B_N]$, and $E_N = [E_1,...,E_N]$ are the observations at a passive eavesdropper (Eve), then the secret key rates generated between Alice and Bob conditioned on Eve $R(A; B || E)$ can be bounded by \cite{Maurer},
\begin{equation} \label{MI}
  I(A;B)-I(A;E) \leq R(A; B || E) \leq I(A;B).
\end{equation}
Here, $I(A;B)$ and $I(A;E)$ are the mutual informations (MI) between the received signals of the pairs of Alice-Bob and Alice-Eve, respectively.

In earlier studies, we conducted experimental measurements to assess the attainable SKG rates based on both channel characteristics and SKG parameters \cite{Mayya_Eucnc,jcas,Mayya_Globe_23}. In this work, we consider the received signal strength (RSS) at Alice, Bob, and Eve over wideband frequency selective channels as the source of shared randomness. Our findings highlight the critical importance of users being channel-aware for a secure implementation of the SKG protocol. The susceptibility of SKG to on-the-shoulder eavesdropping attacks is closely tied to the design parameters. Notably, the level of channel decorrelation between legitimate and adversarial users is contingent on the interference patterns of the multipath observed at the eavesdropper, rather than being solely determined by the physical distance between the eavesdropper and the legitimate users. In this work, the secret key rates are upper bounded by the conditional min-entropy in order to ensure the confidentiality of the extracted secret keys.

 The entropy of the channel depends on the stochastic properties of the channel distribution. This necessitates the need to understand the channel distribution of the given environment in order to estimate the MI and evaluate the theoretical upper bounds for the SKG rates given by \eqref{MI}. 
 This helps in optimally selecting as well as adapting the SKG parameters during a given instance of SKG from the channel measurements.  Hence, in this work, we aim to theoretically characterize the multipath channel models under the line-of-sight scenarios when observed at the receiver in order to evaluate the SKG rates. This is crucial in order to scrutinize the optimality of the practical evaluations. In this context, the focus of this work will be to evaluate how the bandwidth, delay spread, and the Rician $K$-factor impact the MI estimated based on the measured RSS. 
 
 The rest of the paper is organized as follows. In Section~\ref{Channel Model} we describe the system model, and in Section~\ref{RSS_Distributions} we evaluate the distributions of the received signal strength based on how the multipath components (MPC) resolve at the receiver. In Section~\ref{Numerical Estimation} we estimate the upper bounds of the SKG rates given by \eqref{MI}. Finally, Section~\ref{Conclusion} concludes the paper. 

\section{System Model}\label{Channel Model}
\begin{figure}[!t]
    \centering
    \includegraphics[width=0.48\textwidth]{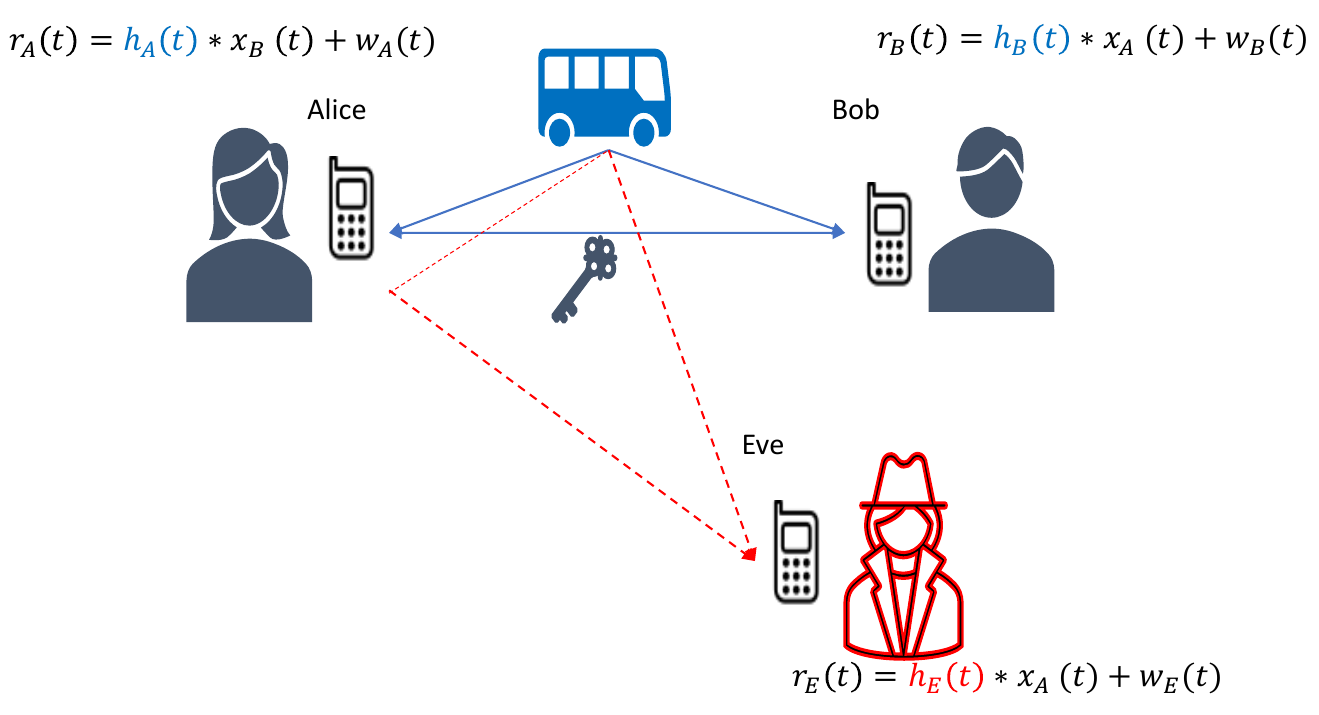}
    \caption{System model with Alice, Bob, and Eve.}   
    \label{fig:system_model}
\end{figure}

The system model is shown in Fig.~\ref{fig:system_model}. 
Alice and Bob exchange complex chirps with constant modulation in a time division duplex (TDD) manner. The power spectral density of a chirp is nearly flat and is well defined. Eve constantly eavesdrops upon this exchange. 
The received signals at Alice, Bob, and Eve can be represented as
\begin{equation}
    r_{l}(t) = x(t) * h_{l}(t) + w_{l}(t),\\
\end{equation}
where $l \in \{A, B, E\}$ stands for Alice, Bob, and Eve, respectively, $x(t)$ is the transmit signal, $w_{l}(t)$ is an additive white Gaussian noise (AWGN) variable. The channel impulse response (CIR) can be written as\cite{Goldsmith}
\begin{equation}
    h(t) = \sum_{n=0}^{L-1} \alpha_n(t)e^{-i\phi_n(t)}\delta(\tau-\tau_n(t)).
\end{equation}

Due to reciprocity between Alice and Bob, $h_A(t) \approx h_B(t) \neq h_E(t)$. Here, $\alpha_n(t)$ is amplitude attenuation, $e^{-i\phi_n(t)}$ is the phase shift, and $\tau_n(t)$ is the time delay of the $n^{th}$ MPC, $L$ is the total number of MPCs. 
Here, $\alpha_i, i = 0,...,N-1 $ are complex Gaussian variables \cite{3gpp}. The magnitudes $\lvert\alpha_i\rvert$ are assumed to follow a $Rayleigh$ distribution for non-line of sight (NLoS) channels and a $Rician$ distribution for LoS channels. In this work, we assume LoS channel models with independently distributed MPCs.

The multipath scenario with $L$ paths can be represented as one LoS path and $L-1$ diffuse paths. The underlying complex Gaussian RVs of the LoS path is modelled as $Z = X_1+iY_1$, where $X$ and $Y$ are identical and independently distributed (i.i.d.) and can be represented as 
$X_1\sim\mathcal{N}(\nu\cos\theta, \sigma^2)$ and $Y_1\sim\mathcal{N}(\nu\sin\theta, \sigma^2)$, respectively. The diffuse paths are also complex Gaussian RVs $Z_i = X_i+iY_i$, $i =2,...,L$, where $X_i$ and $Y_i$ are i.i.d. with $X_i\sim\mathcal{N}(0, \sigma^2)$ and $Y_i\sim\mathcal{N}(0, \sigma^2)$, respectively. The $Rician$ distribution parameter $K$-factor is described as the ratio between the power in the direct path and the power in other paths.


To exploit the frequency selectiveness of the channel, we extract the shared channel randomness from the multipath in the frequency domain. The received signals are filtered frame-wise using a unit gain low pass filter with cut-off frequency $B_w/2$, where $B_w$ is the filter bandwidth (BW). The measured RSS is given by
\begin{equation}\label{RSS}
    \hat{P}_{l} = \overline{\lvert g(t)*y_l(t)\rvert^2},
\end{equation} 
where the operator $\overline{(\cdot)}$ denotes time averaging and $g(t)$ is the low pass filter impulse response.
\section{Received Signal Strength Distribution}\label{RSS_Distributions}

The RSS in \eqref{RSS} denotes the received power measurements. Due to the almost flat spectrum of the transmit signal and the low pass filter, the changes observed in the RSS is due to the frequency selective channel attenuation components and the receiver noise. These both are treated as independent processes in order to evaluate the power distributions of the channel attenuations and noise separately and then added together to obtain the distribution of the received power distribution. Here, we evaluate the distribution of Alice, i.e., $\hat{P_A} = [\hat{P_{A1}}, ..., \hat{P_{AN}}]$. The distributions of Bob and Eve will be similar considering the LoS channel model.

 \subsection{Channel Power Distribution When Multipaths Cannot be Resolved}\label{sec:case1}

If $T_m < T$, where $T$ is the signal period and $T_m$ is the multipath delay spread, then all multipath components arrive almost at the same time at the receiver and cannot be individually resolved. This can be observed in narrowband systems. For the multipath with $L$ paths, resolving the multipath components, the final amplitude follows a Rician distribution $R \sim Rice(\nu, L\sigma^2)$. Normalizing the variance, $\frac{R^2}{L\sigma^2}$ is non-central chi square with non-centrality parameter $\frac{\nu^2}{L\sigma^2}$ and 2 degrees of freedom, i.e., 
\begin{equation}\label{eq:ncx2_1}
    \frac{1}{\sigma^2}Z^2\sim\chi^2\left(2, \frac{\nu^2}{L\sigma^2}\right).
\end{equation}

\subsection{Channel Power Distribution When All Multipaths Resolve Separately}\label{sec:case2}
If $T_m > T$, the multipath components are resolved individually. This causes wideband fading and are observed in narrowband channel models. For the multipath with $L$ paths, resolving the multipath components, the final amplitude follows a Rician distribution $R \sim Rice(\nu, \sigma^2)$.


Normalizing the variance, $\frac{R^2}{\sigma^2}$ is non-central chi square with non-centrality parameter $\frac{\nu^2}{\sigma^2}$ and $2L$ degrees of freedom, i.e., 
\begin{equation}\label{eq:ncx2_2}
    \frac{1}{\sigma^2}Z^2\sim\chi^2\left(2L, \frac{\nu^2}{\sigma^2}\right).
\end{equation}


 \subsection{Channel Power Distribution for Hybrid Resolution of Multipaths}\label{sec:case3}
Let us have $M$ resolved delay bins with non-resolved MPCs of each delay bin $L_1, L_2, ... L_M$. Then the received power is
\begin{equation}\label{eqn:ch_dist_3}
     Z^2 = h(t)^2 = \sum_{i_1=0}^{L_1-1}\alpha_{i_1}^2 + \sum_{i_2=0}^{L_2-1}\alpha_{i_2}^2 + \hdots +\sum_{i_M=0}^{L_M-1}\alpha_{i_M}^2.
\end{equation}
The number of resolvable delay bins is given by $M = \ceil*{T_m/T}$, where $\ceil*{.}$ corresponds to the ceiling operator.
The power of first delay bin $L_1$ consists of the LoS component or the direct path. Hence, 
\begin{equation} \label{eqn:11}
    \sum_{i_1=0}^{L_1-1}\alpha_{i_1}^2 \sim\chi^2\left(2, \frac{\nu^2}{L_1\sigma^2}\right).
\end{equation}

The power of remaining delay bins $L_2, L_3, ..., L_M$ is

\begin{equation}\label{eqn:15}
\begin{split}
    \sum_{i_2=0}^{L_2-1}\alpha_{i_2}^2 + \sum_{i_3=0}^{L_3-1}\alpha_{i_3}^2+ \hdots +\sum_{i_M=0}^{L_M-1}\alpha_{i_M}^2 \sim \qquad\qquad\quad \\
    \Gamma \left(\frac{2*\sum_{i=2}^M 2*L_i*\sigma^2}{(\sum_{i=2}^M 2*L_i*\sigma^2)^2}, \frac{\sum_{i=2}^{M}(2*L_i*\sigma^2)^2}{\sum_{i=2}^M2*L_i*\sigma^2}\right).
\end{split}
\end{equation}
Here, we use the Welch-Satterwaite approximation \cite{MITEV_vtc2022}. The channel distribution is the sum of the distributions in \eqref{eqn:11} and \eqref{eqn:15} and its probability distribution function (pdf) is obtained by the convolution of the two pdfs.
Thus,
\begin{equation}\label{eqn:case3}
\begin{split}
    \frac{1}{\sigma^2}Z^2 \sim  \chi^2\left(2, \frac{\nu^2}{L_1\sigma^2}\right) + \quad\\ \Gamma \left(\frac{2*\sum_{i=2}^M 2*L_i*\sigma^2}{(\sum_{i=2}^M 2*L_i*\sigma^2)^2}, \frac{\sum_{i=2}^{M}(2*L_i*\sigma^2)^2}{\sum_{i=2}^M2*L_i*\sigma^2}\right).
\end{split}
\end{equation}

\subsection{Noise Power Distribution}
The noise power distribution is found to be dependent on the number of samples which directly depends on the sampling rate at the receiver and the frame resolution. For a frame of $C$ samples, the noise power is distributed as \cite{MITEV_vtc2022}
\begin{equation} \label{eq:noise_dist}
    N_C^2 \sim \Gamma\left(\frac{C-1}{2}, \frac{4\sigma^2}{C-1}\right).
\end{equation}

\subsection{Total Power Distribution}
The total power distribution of the RSS is the sum of the corresponding channel power distributions and the noise power distribution, whose pdfs can be obtained by convolution.

\begin{figure*}
\centering
\minipage[t]{\linewidth}
    \includegraphics[height = 0.30\textheight,width=\textwidth]{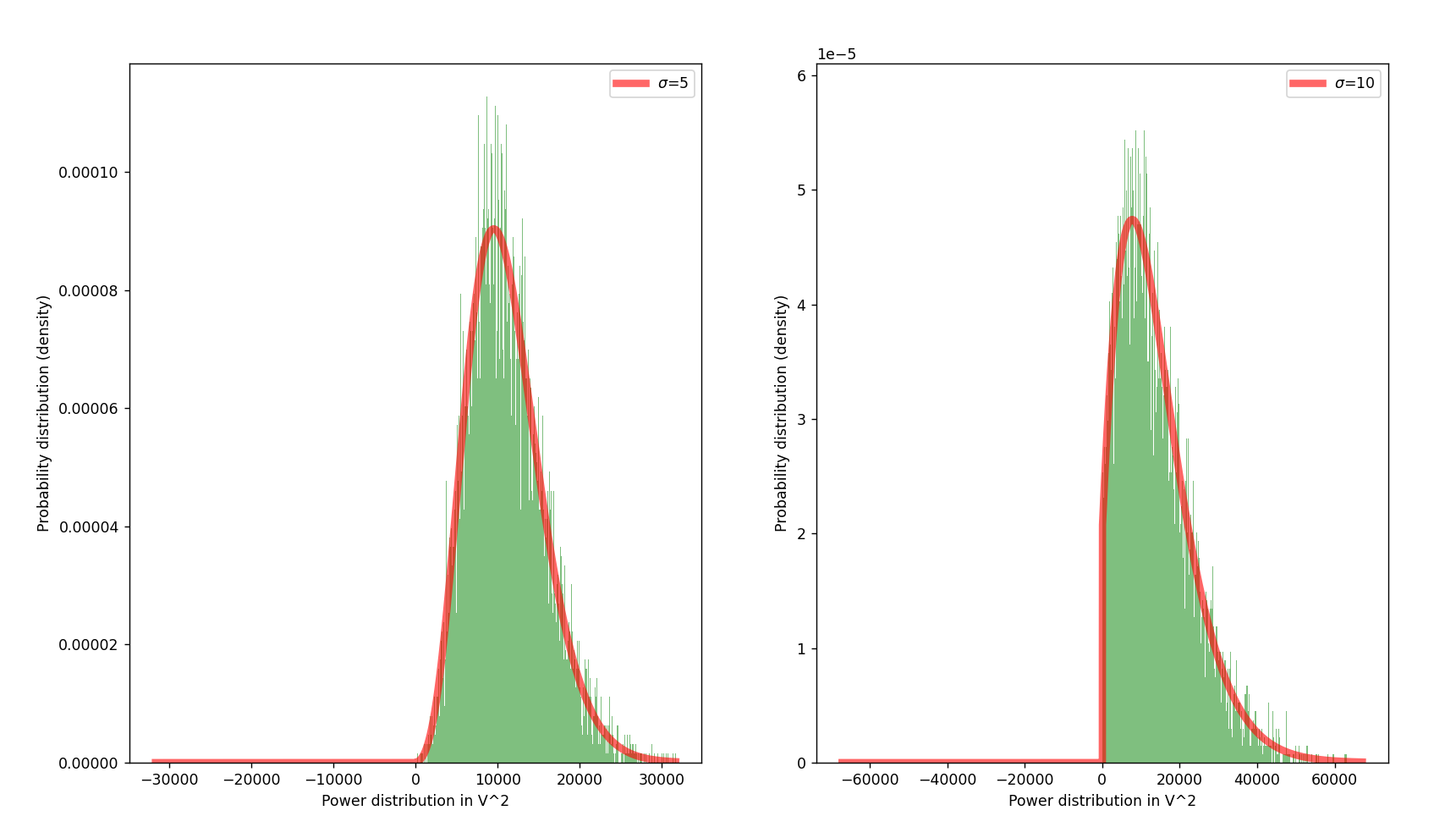}
    \caption{Total power distribution when multipaths cannot be resolved. The number of MPCs is $L=14$ and the underlying Gaussian RVs are generated for $\sigma = \{5, 10\}$ and $\nu=100$.}
\label{fig:Case1}
    \includegraphics[height = 0.30\textheight,width=\textwidth]{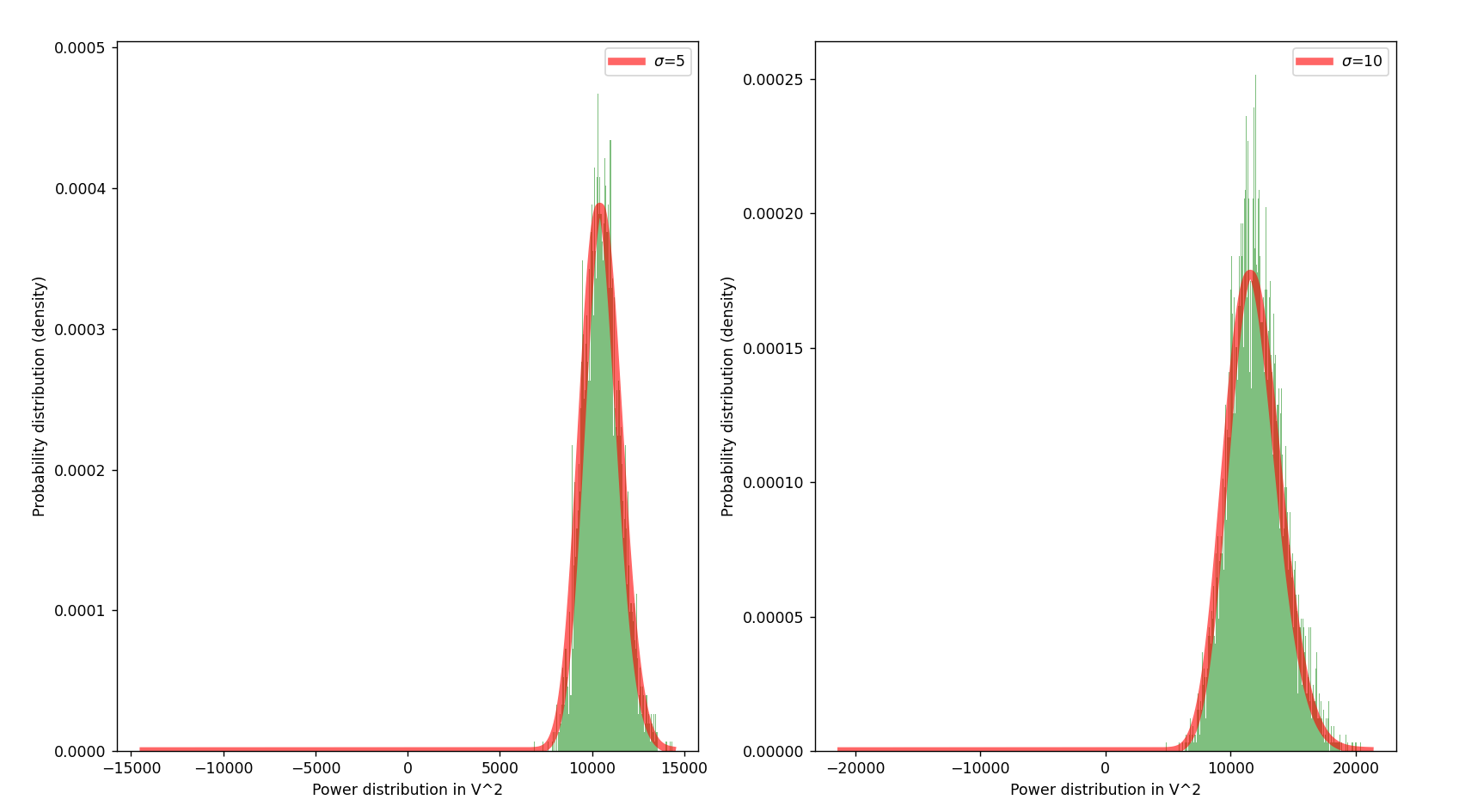}
    \caption{Total power distribution when all multipaths resolve separately. The number of MPCs is $L=14$ and the underlying Gaussian RVs are generated for $\sigma = \{5, 10\}$ and $\nu=100$.}
    \label{fig:Case2}

    \includegraphics[height = 0.30\textheight,width=\textwidth]{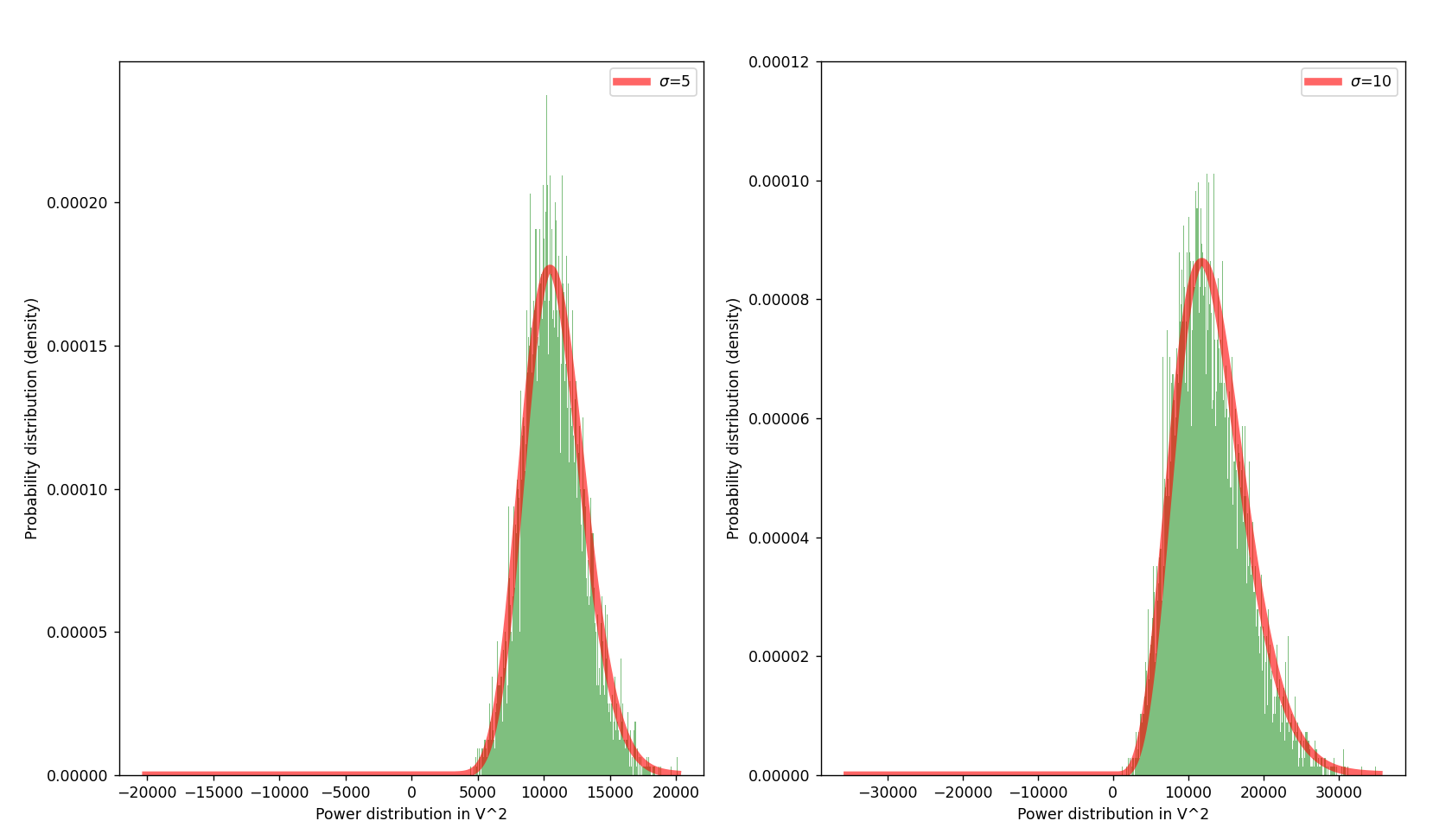}
    \caption{Total power distribution for hybrid resolution of multipaths. The number of MPCs is $L = 14$ and the underlying RVs are generated for $\sigma = \{5, 10\}$, $\nu=100$, $M = 4$, $L_1 = 3$, $L_2 = 5$, $L_3 = 4$, and $L_4 =2$ (see also \eqref{eqn:ch_dist_3}).}
    \label{fig:Case3}
\endminipage
\end{figure*}

Fig.~\ref{fig:Case1}, Fig.~\ref{fig:Case2}, and Fig.~\ref{fig:Case3} show the total power distributions (RSS distributions) for the three cases of the channel resolutions represented above in Section~\ref{sec:case1}, Section~\ref{sec:case2}, and Section~\ref{sec:case3}, respectively. Here, the pdfs of corresponding channel distributions expressed in \eqref{eq:ncx2_1}, \eqref{eq:ncx2_2}, and \eqref{eqn:case3}, respectively, are convolved with the pdf of the noise distribution in \eqref{eq:noise_dist} to obtain the total power distribution. For all the cases, the distribution parameters are the number of MPCs $L = 14$. The parameters for the Gaussian RVs are generated with $\nu = 100$ and $\sigma = \{5, 10\}$. The parameters in \eqref{eqn:ch_dist_3} are considered as $M = 4$, $L_1 = 3$, $L_2 = 5$, $L_3 = 4$, and $L_4 =2$.

\subsection{Achievable SKG Rates}
As explained before, the secret key rates of the channel are bounded by \eqref{MI}. The MI between the RVs $(A,B)$ is given by
\begin{equation}\label{eqn:12}
  I(A;B) =  \int_{\mathcal B} \int_{\mathcal A}
      {P_{(A,B)}(a,b) \log{ \left(\frac{P_{(A,B)}(a,b)}{P_A(a)\,P_B(b)} \right) }
  } \; \mathrm{d}a \,\mathrm{d}b.
\end{equation}

In the previous sections, we have obtained the expressions for the marginal pdfs $P_A(a)$, $P_B(b)$ (We assume the same pdf of Eve $P_E(e)$). However, obtaining the joint pdfs $P_{(A,B)}(a,b)$ or $P_{(A,E)}(a,e)$ and solving the integral is out of the scope of this work. Hence, we rely on numerical estimators to evaluate the MI. 

The MI of Eve w.r.t. Alice and Bob, respectively, strongly relies on the correlations in the observations. Since we use the channel models to do numerical estimations, we cannot realistically consider this parameter. For the sake of brevity, we evaluate the upper bounds of the MI in \eqref{MI} and leave the evaluation of the lower bounds for future work.

\section{Numerical Estimation} \label{Numerical Estimation}
In this section we evaluate the MI numerically as denoted in \eqref{eqn:12}. For this we use the local non-uniformity correction (LNC) estimator which estimates the MI with local non-uniformity correction \cite{LNC} as it has proven to be reliable even at low signal-to-noise ratio (SNR) values compared to other estimators. For this work, we use the 3GPP tapped delay line~E (TDL-E) model which corresponds to LoS channel profiles \cite{3gpp}. This has 14 MPCs with the first tap corresponding to LoS, i.e., Rician distribution, and the remaining taps corresponding to the Rayleigh distribution as assumed in the previous section. The transmit waveform $x(t)$ is a complex chirp signal with constant modulation and BW of $250$ MHz. This waveform is convolved with the CIRs $h_l(t)$ generated according to the TDL-E model and AWGN noise $w_l(t)$ is added to simulate the received signal $y_l(t)$ for Alice and Bob (Eve). The received signal powers are then convolved using a low pass filter $g(t)$ with cut-off frequency $B_w/2$ and the square magnitude is averaged over the filter bandwidth $B_w$. These power values $P_l$ are then used in the SKG protocol to derive the keys. In this work, based on different MIs, delay spread, and $K$-factor, we simulate the received signals and the power values. We aim to see the impact of these parameters on the MI evaluated. These power values, $P_l$ are congruent to the RVs whose theoretical distributions is derived in the Section~\ref{RSS_Distributions}.

\begin{figure}
    \centering
    \includegraphics[width=0.48\textwidth]{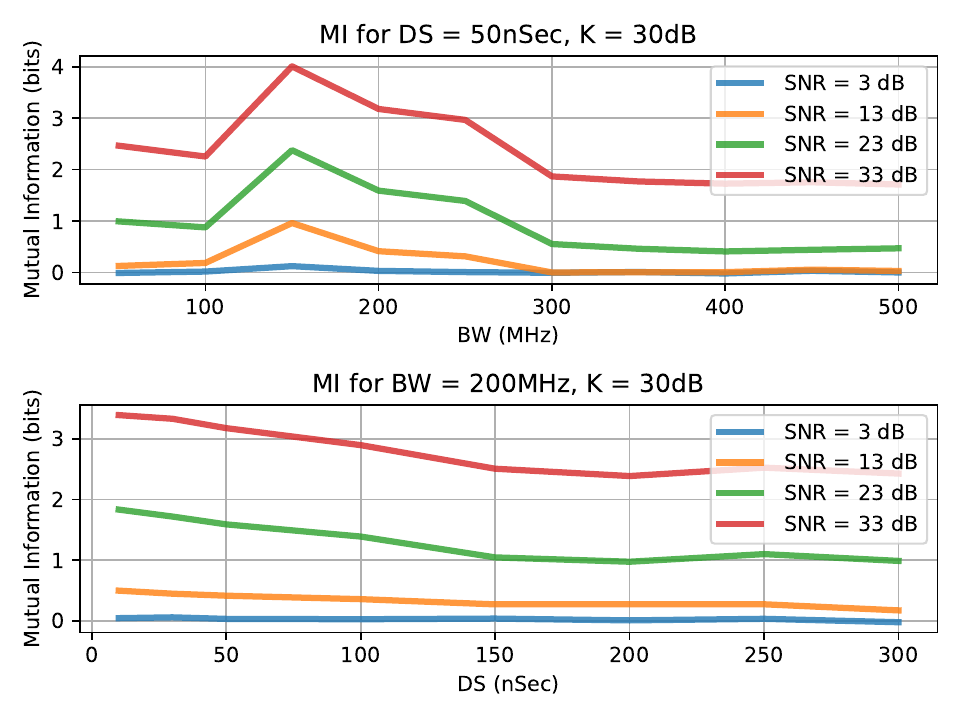}
    \caption{MI evaluated at constant K = 30dB for \textbf{top}: different SNRs for constant DS = 50ns and varying BW, \textbf{bottom}:  different SNRs for constant BW = 200MHz and varying DS. }   
    \label{fig:MI_BW_DS}
\end{figure}

Fig.~\ref{fig:MI_BW_DS} shows dependence of MI values on the SNRs. The top figure depicts the MI for different MIs for constant delay spread of $50$ ns and constant $K$-factor of $30$ dB. Here, we note that, as the bandwidth increases, the MI decreases. At higher bandwidths, more number of MPCs can be resolved at the receiver, this leads to decrease in the variance which causes the MI to decrease. After $300$ MHz, since the MI remains constant with increasing BW, we conclude that all the MPCs have resolved in different delay bins leading to constant variance which decreases the randomness and the MI observed.

The figure at the bottom represents how the different delay spreads impact the MI estimated for the given bandwidth of $200$ MHz and $K$-factor of $30$ dB. Here again we note that, for lower delay spreads, we have higher MI. The MPCs fall into the same delay bin at low delay spreads meaning they are unresolvable. This leads to higher variance in the resulting distribution increasing the MI. After $150$ ns, the MI remains constant for further increase in the DS which means all the MPCs are resolved leading to constant variance due to which the randomness and the estimated MI decrease.

\begin{figure}
    \centering
    \includegraphics[width=0.48\textwidth]{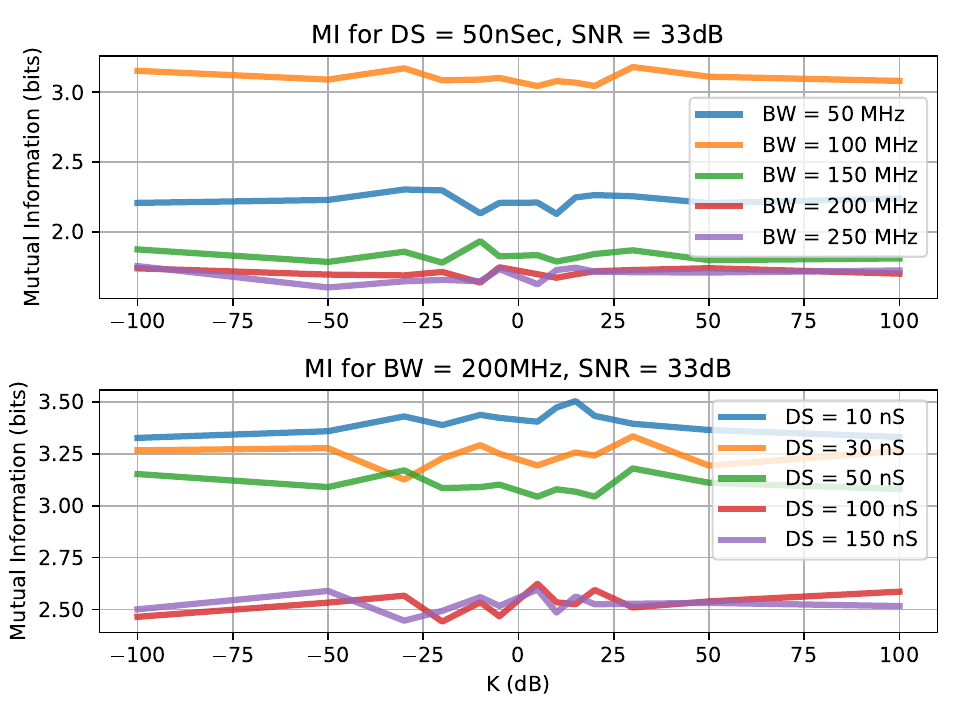}
    \caption{MI evaluated at constant SNR = 33dB and different K-factors for \textbf{top}: constant DS = 50ns  \textbf{bottom}: constant BW = 200MHz. }   
    \label{fig:MI_BW_DS_K}
\end{figure}

As the SNR increases, the evaluated MI also increases as expected. At lower SNRs, the signal power will be only the power of the direct path of the LoS channel which is expected to be the same for both Alice and Bob, effectively decreasing the randomness further. Hence, for the analysis of the impact of $K$-factor on the MI in the Fig.~\ref{fig:MI_BW_DS_K}, we consider the SNR of $33$ dB.

Fig.~\ref{fig:MI_BW_DS_K} shows the effects of $K$-factor on the MI. In the top figure, the MI due to different BWs at constant DS of $50$ ns is shown. Here, similar to Fig.~\ref{fig:MI_BW_DS} we see that increasing the BW decreases the MI, however increasing the $K$-factor does not have a greater impact on the evaluated MI. Similar conclusion can be drawn for the bottom figure which represents the MI due to different DSs at constant BW of $200$ MHz. Increasing the DS decreases the MI, but for a given DS, the MI remains constant across all $K$-factors. This can be attributed to the reciprocity in Alice and Bob's observations. Due to this reciprocity, Alice and Bob observe the similar $K$-factor which means it can be considered as a shared information which is not random and hence does not contribute to the MI extracted.

From these graphs, we conclude that, the MI which serves as the upper bound for the evaluated SKG rate depends on the resolution BW of the filterbank, the delay spreads and the SNR. How the MPCs resolve at the receiver also impacts the variance of the distribution and hence the achievable MI. \textit{This signifies the fact that in order to implement the SKG and evaluate the SKG rates, the users must be channel-aware}.

\section{Conclusion}\label{Conclusion}
In this work, we analyzed how the channel characteristics of the Rician channel impact the achievable MI and hence the achievable SKG rates. To do this, we evaluated the channel distributions of the RSS depending on how the MPCs resolve. This depends on the bandwidth and the delay spread of the filterbank at the receiver. We then numerically evaluated the MI for TDL-E channel model for varying bandwidths, delay spread, and $K$-factors. This work shows that the resolvability of the MPCs at the receiver has an impact on the MI, i.e., if all MPCs fall into the same delay bin, this leads to increased variance in the distribution and hence increased MI, while if the MPCs resolve separately into different delay bins, the variance decreases and the evaluated MI also decreases. The MI decreases with increased BW and DS while the $K$-factor itself does not have an impact on the evaluated MI. In future work, we aim to extend this work to experimental datasets and to evaluate how the eavesdroppers correlation impacts the lower bounds of the SKG rates.
\balance
\bibliographystyle{IEEEtran}
\bibliography{bib}

\begin{thebibliography}{10}
\providecommand{\url}[1]{#1}
\csname url@samestyle\endcsname
\providecommand{\newblock}{\relax}
\providecommand{\bibinfo}[2]{#2}
\providecommand{\BIBentrySTDinterwordspacing}{\spaceskip=0pt\relax}
\providecommand{\BIBentryALTinterwordstretchfactor}{4}
\providecommand{\BIBentryALTinterwordspacing}{\spaceskip=\fontdimen2\font plus
\BIBentryALTinterwordstretchfactor\fontdimen3\font minus \fontdimen4\font\relax}
\providecommand{\BIBforeignlanguage}[2]{{%
\expandafter\ifx\csname l@#1\endcsname\relax
\typeout{** WARNING: IEEEtran.bst: No hyphenation pattern has been}%
\typeout{** loaded for the language `#1'. Using the pattern for}%
\typeout{** the default language instead.}%
\else
\language=\csname l@#1\endcsname
\fi
#2}}
\providecommand{\BIBdecl}{\relax}
\BIBdecl

\bibitem{jsac}
Y.~Wu, A.~Khisti, C.~Xiao, G.~Caire, K.-K. Wong, and X.~Gao, ``A survey of physical layer security techniques for {5G} wireless networks and challenges ahead,'' \emph{IEEE J. Sel. Areas Commun.}, vol.~36, no.~4, pp. 679--695, 2018.

\bibitem{chorti2022contextaware}
A.~Chorti, A.~N. Barreto, S.~Köpsell, M.~Zoli, M.~Chafii, P.~Sehier, G.~Fettweis, and H.~V. Poor, ``Context-aware security for {6G} wireless: The role of physical layer security,'' \emph{IEEE Commun. Stand. Mag.}, vol.~6, no.~1, pp. 102--108, 2022.

\bibitem{plsi}
R.~F. Schaefer and H.~Boche, ``Physical layer service integration in wireless networks: Signal processing challenges,'' \emph{IEEE Signal Process. Mag.}, vol.~31, no.~3, pp. 147--156, 2014.

\bibitem{BlochBarros-2011-PhysicalLayerSecurity}
M.~Bloch and J.~Barros, \emph{Physical-Layer Security: From Information Theory to Security Engineering}.\hskip 1em plus 0.5em minus 0.4em\relax Cambridge, UK: Cambridge University Press, 2011.

\bibitem{SchaeferBocheKhistiPoor-2017-InformationTheoreticSecurityPrivacy}
R.~F. Schaefer, H.~Boche, A.~Khisti, and H.~V. Poor, Eds., \emph{Information Theoretic Security and Privacy of Information Systems}.\hskip 1em plus 0.5em minus 0.4em\relax Cambridge, UK: Cambridge University Press, 2017.

\bibitem{pnas}
H.~V. Poor and R.~F. Schaefer, ``Wireless physical layer security,'' \emph{Proc. Natl. Acad. Sci. U.S.A.}, vol. 114, no.~1, pp. 19--26, Jan. 2017.

\bibitem{Maurer}
U.~M. Maurer, ``Secret key agreement by public discussion from common information,'' \emph{IEEE Trans. Inf. Theory}, vol.~39, no.~3, pp. 733--742, 1993.

\bibitem{ahlswede}
R.~Ahlswede and I.~Csiszar, ``Common randomness in information theory and cryptography. i. secret sharing,'' \emph{IEEE Trans. Inf. Theory}, vol.~39, no.~4, pp. 1121--1132, 1993.

\bibitem{reciprocity_Smith}
G.~Smith, ``A direct derivation of a single-antenna reciprocity relation for the time domain,'' \emph{IEEE Trans. Antennas Propag.}, vol.~52, no.~6, pp. 1568--1577, 2004.

\bibitem{half_wavelength}
W.~{Jakes}, \emph{Microwave Mobile Communications}.\hskip 1em plus 0.5em minus 0.4em\relax Wiley-IEEE Press, 1994.

\bibitem{Goldsmith}
A.~Goldsmith, \emph{Wireless Communications}.\hskip 1em plus 0.5em minus 0.4em\relax USA: Cambridge University Press, 2005.

\bibitem{He2016linksignature}
X.~He, H.~Dai, W.~Shen, P.~Ning, and R.~Dutta, ``Toward proper guard zones for link signature,'' \emph{IEEE Trans. Wireless Commun.}, vol.~15, no.~3, pp. 2104--2117, 2016.

\bibitem{Mayya_Eucnc}
A.~Mayya, M.~Mitev, A.~Chorti, and G.~Fettweis, ``Effects of channel characteristics and design parameters on secret key generation rates,'' in \emph{European Conf. Networks and Communications}, Gothenburg, Sweden, Jun. 2023, pp. 1--2.

\bibitem{jcas}
M.~Mitev, A.~Mayya, and A.~Chorti, ``Joint secure communication and sensing in {6G} networks,'' in \emph{Physical Layer Security for 6G}.\hskip 1em plus 0.5em minus 0.4em\relax Wiley-IEEE Press, 2024.

\bibitem{Mayya_Globe_23}
A.~Mayya, M.~Mitev, A.~Chorti, and G.~Fettweis, ``A {SKG} security challenge: Indoor {SKG} under an on-the-shoulder eavesdropping attack,'' in \emph{IEEE Global Commun. Conf.}, Kuala Lumpur, Malaysia, Dec. 2023, pp. 1--6.

\bibitem{3gpp}
``{3{GPP}, Release 16, TR 38.901, Study on channel model for frequencies from 0.5 to 100 GHz}.''

\bibitem{MITEV_vtc2022}
M.~Mitev, A.~N. Barreto, T.~M. Pham, and G.~Fettweis, ``Secret key generation rates over frequency selective channels,'' in \emph{IEEE 95th Veh. Tech. Conf (VTC2022-Spring)}, Helsinki, Finland, Jun. 2022, pp. 1--5.

\bibitem{LNC}
S.~Gao, G.~Ver~Steeg, and A.~Galstyan, ``Efficient estimation of mutual information for strongly dependent variables,'' in \emph{Proc. Eighteenth Int. Conf. Artificial Intelligence and Statistics}, ser. Proceedings of Machine Learning Research, G.~Lebanon and S.~V.~N. Vishwanathan, Eds., vol.~38.\hskip 1em plus 0.5em minus 0.4em\relax San Diego, California, USA: PMLR, 09--12 May 2015, pp. 277--286.

\end{thebibliography}

\end{document}